\documentclass{emulateapj}

\bibpunct[, ]{(}{)}{;}{a}{}{,}

\begin{document}

   \title 
   {The spectral lag of GRB\,060505: a likely member of the long duration class}

\shorttitle{The spectral lag of GRB\,060505}
\shortauthors{McBreen et al.}

   \author{ 
         S.~McBreen\altaffilmark{1},
                    S.~Foley\altaffilmark{2},
	   D.~Watson\altaffilmark{3},
	    L.~Hanlon\altaffilmark{2},
	    	   D.~Malesani\altaffilmark{3},
	   	   J.~P.~U.~Fynbo\altaffilmark{3},
	   D.~A.~Kann\altaffilmark{4},
	   N.~Gehrels\altaffilmark{5},
	   S.~McGlynn\altaffilmark{2},
	     D.~Palmer\altaffilmark{6} }
         
   \altaffiltext{1}{Max-Planck-Institut f\"{u}r extraterrestrische Physik, 85748 Garching, Germany; smcbreen@mpe.mpg.de}
      \altaffiltext{2}{School of Physics, University College Dublin, Dublin 4, Ireland; sfoley, smcglynn@bermuda.ucd.ie, lorraine.hanlon@ucd.ie} 
         \altaffiltext{3}{Dark Cosmology Centre, Niels Bohr Institute, University of Copenhagen, Juliane Maries Vej 30, DK-2100 Copenhagen \O, Denmark; darach, malesani, jfynbo@dark-cosmology.dk} 
            \altaffiltext{4}{Th\"uringer Landessternwarte Tautenburg, Sternwarte 5,  D-07778 Tautenburg, Germany; kann@tls-tautenburg.de} 
    \altaffiltext{5}{NASA/Goddard Space Flight Center, Greenbelt, Maryland 20771, USA; gehrels@milkyway.gsfc.nasa.gov} 
\altaffiltext{6}{Los Alamos National Laboratory, Los Alamos, New Mexico 87545, USA;  palmer@lanl.gov
}

   \begin{abstract}
   Two long $\gamma$-ray bursts, 
GRB\,060505 and  GRB\,060614,     occurred in nearby galaxies at
    redshifts of 0.089 and 0.125 respectively. Due to their proximity and durations, 
    deep follow-up campaigns to search for 
    supernovae (SNe) were initiated. 
        However none were found in either case, to limits more than
    two orders of magnitude fainter than the prototypical GRB-associated SN,
    1998bw. It was suggested that the bursts, in spite of their durations
    ($\sim4$ and 102\,s), belonged to the population of
    \textit{short} GRBs 
    which has been shown  to be unrelated to SNe. 
        In the case of GRB\,060614 this
    argument was based on  a number of indicators, including the negligible spectral lag,  
    which is consistent with that of short
    bursts.  GRB\,060505 has a shorter duration, but no spectral lag was
    measured.  We present the spectral lag measurements of GRB\,060505
    using $\emph{Suzaku}$'s Wide Area Monitor and the $\emph{Swift}$ Burst
    Alert Telescope.
     We find that the lag is  $0.36\pm0.05$\,s, inconsistent with the lags of short bursts and
    consistent with the properties of long bursts and SN-GRBs. 
          These results  support the association of GRB\,060505 with
    other low-luminosity GRBs also found in star-forming galaxies
       and indicates that 
    at     least 
    some massive stars may die without bright SNe.
   \end{abstract}


%
%

\section{Introduction\label{introduction}}

The existence of two classes of gamma-ray bursts (GRBs) differing in
observed durations and spectral properties has been established for some
time
\citep[e.g.][]{1981Ap&SS..80..119M,1984Natur.308..434N,1992AIPC..265....3H}.
These populations were quantified using the Burst and Transient Source
Experiment (BATSE), which showed a  bimodal distribution in the
durations of GRBs well fit by two lognormal functions
\citep{1994MNRAS.271..662M}, with the divide at $\sim2\,$s
\citep{1993ApJ...413L.101K}.   
In addition,  there is also contamination in the short burst class from soft gamma-ray repearters \citep[e.g.][]{2008arXiv0802.0008C}. 
It is generally accepted that long GRBs
have their origins in massive star progenitors because of their  association with core-collapse supernovae  
\citep[SNe,][]{1998Natur.395..670G,2003Natur.423..847H,2003ApJ...591L..17S,2004ApJ...609L...5M,2006Natur.442.1011P,2006ARA&A..44..507W}
 and
 occurrence in star-forming galaxies \citep{2002AJ....123.1111B} and in
highly star-forming regions therein \citep{2006Natur.441..463F}.
The origin of short GRBs is still open, with mergers of compact
objects being the leading concept
\citep[e.g.][]{2005Natur.437..851G,2005Natur.437..859H,2005Natur.437..845F}.

The detection of 
the spectroscopic signatures of SNe in the 4 nearest GRBs and the detection of
bumps consistent with SNe in the lightcurves of most low-redshift bursts
seemed to confirm the paradigm that all long GRBs would be associated with
SNe \citep{2004ApJ...609..952Z,2006ARA&A..44..507W}, as predicted by the
 collapsar model of long GRBs \citep{1999A&AS..138..499W}.
Doubts were cast on 
this paradigm 
by the non-detection of SNe in
two nearby GRBs, GRB\,060505 at $z=0.089$ \citep{2006GCN..5123....1O}
and
GRB\,060614 at $z=0.125$ \citep{Price06_GCN5275}
discovered by the \emph{Swift} \citep{gehrels:2004} Burst Alert Telescope
\citep[BAT,][]{2005SSRv..120..143B}. Due to their long durations, $T_{90}$ of $4\pm1$\,s and $102\pm5$\,s respectively 
\citep{2006GCN..5142....1H,Barthelmy06_GCN5256}, 
SN searches were initiated. 
Although a supernova $\sim$100
times fainter than SN1998bw would have been detected, none was found in
either case
\citep{2006Natur.444.1047F,2006Natur.444.1053G,2006Natur.444.1050D,2007ApJ...662.1129O}. 
It was suggested that they were short bursts where 
the lack of SNe  would not be surprising, as short GRBs
have not  shown 
 SN emission 
 \citep{2005Natur.437..859H,2005Natur.437..845F,2006ApJ...638..354B,2006A&A...447L...5C}.

The classification of GRBs with durations close to the
long/short  division is problematic. The argument that GRB\,060614 was a
``short GRB" rests  on its extended soft emission component
and  on its negligible spectral lag 
\citep{2006Natur.444.1044G,2007ApJ...655L..25Z}. When the latter is combined with its
relatively low luminosity, it violates the lag-luminosity relation found by
\citet{2000ApJ...534..248N} for long GRBs. 
If the lack of a SN in GRB\,060505 is to be attributed to it being a short burst, it should also have a negligible lag. 
We present the spectral lag analysis of the prompt emission of
GRB\,060505 using data from the \emph{Suzaku}  Wide Area
Monitor  (WAM) and \emph{Swift}-BAT.

\section{Observations and data reduction}

\begin{figure}[t]
\begin{centering}
\epsscale{1}
\plotone{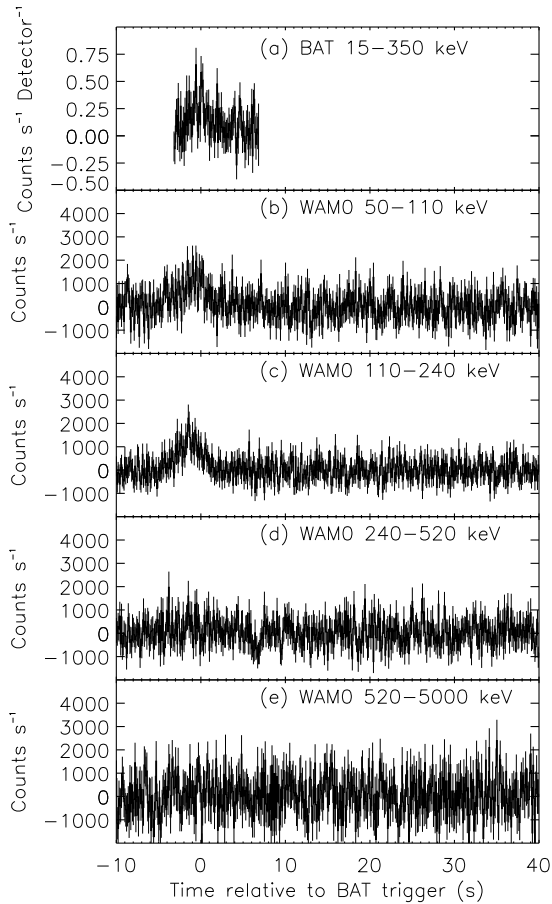}
 \caption{The lightcurves of GRB\,060505 with the BAT instrument on
         $\emph{Swift}$ (a) and from the WAM0 detector on $\emph{Suzaku}$ (b-e), all at 100\,ms resolution. 
                  Time is relative to the BAT trigger. 
                 A precursor is visible in the lowest energy WAM channel.}
\label{fig:BATlightcurves}\label{fig:WAWlightcurves}
\end{centering}
\end{figure}

GRB\,060505 was detected by the BAT instrument on \emph{Swift}.  
The fluence  
is  (6.2$\pm$1.1)$\times 10^{-7}$ ergs cm$^{-2}$ (15-150\,keV) and the spectrum is fit by a power law with index 1.3$\pm0.3$ 
\citep{2006GCN..5142....1H}. The trigger fell below the
6.5$\sigma$ threshold for an automatic slew but 
 ground analysis found a   8.5$\sigma$ excess \citep{2006GCN..5076....1P}.
\emph{Swift} was
repointed at T$_0$+0.6~days and a weak fading X-ray source was identified
\citep{2006GCN..5114....1C}. 
We obtained the publicly available data  for GRB\,060505    from
the $\emph{Swift}$
archive\footnote{http://swift.gsfc.nasa.gov/docs/swift/archive/}. A mask
weighted lightcurve was generated using the  BAT data analysis
tools. The available data contained only 10\,s of event data and the
lightcurve is presented in Fig.~\ref{fig:BATlightcurves}.  
$\emph{Swift}$ was approaching the South Atlantic Anomaly when the burst occurred and was subject to a higher than normal background level.  Additionally, the partial
coding was only 11\% \citep{2006GCN..5142....1H} meaning that the off-axis
angle with respect to the $\emph{Swift}$ axis was almost 50$\degr$,
substantially reducing the effective area of the instrument. Splitting the
data into energy channels  for spectral lag analysis 
further
reduces the weak signal.  
 
The WAM is the anti-coincidence shield (ACS) of the Hard X-ray Detector on $\emph{Suzaku}$
\citep{2006SPIE.6266E.122Y} and it also triggered on GRB\,060505.
The WAM consists of four identical walls which act as
individual detectors (WAM0 to WAM3). The detectors have a large effective
area  \citep{2006SPIE.6266E.122Y}. They are sensitive in the
energy range 50--5000\,keV, and although its primary role is to act as
an ACS, WAM is also used as an all-sky monitor for 
 GRBs. An automated triggering system operates on board
\citep{2006AIPC..836..201Y} and the lightcurves are publicly available at
15.6\,ms resolution in 4 rough energy bands 50--110, 110--240, 240--520,
520--5000\,keV\footnote{http://www.astro.isas.ac.jp/suzaku/HXD-WAM/WAM/}.
The lightcurves in the four energy channels from the WAM0 detector
at 100\,ms resolution are presented in Fig.~\ref{fig:WAWlightcurves}. The T$_{\rm 90}$ of
GRB\,060505 was $\sim$4.8~s in the
50--5000\,keV band$^{8}$.
The burst struck the WAM detector at an angle 
such that principally WAM0, but also to a lesser extent WAM3, detected the burst.
The on-axis effective area of the BAT and WAM instruments are shown in Fig.~2 of
\citet{2006SPIE.6266E.122Y} and the effective area of WAM only exceeds that
of BAT above 300\,keV. However, it should be remembered that GRB\,060505
occurred $\sim50\degr$ off-axis in BAT 
and that the effective area of BAT also drops rapidly above 100\,keV.
These factors result in a more significant detection 
GRB\,060505 
by WAM than BAT and therefore we rely primarily on the WAM data for our
analysis. However, we show that the results are consistent with those obtained
from the BAT data.

\section{Data Analysis and Results}

The spectral lag  was calculated by
cross-correlating the lightcurves in different energy channels
\citep[e.g.][]{1997ApJ...486..928B,2000ApJ...534..248N,Foley2007}. The
cross-correlation function (CCF) was fit with a fourth order polynomial and
the quoted lag value is the peak of this function. Statistical errors were
calculated using a bootstrap method as described in
\citet{2000ApJ...534..248N}. 
This involves adding
Poissonian noise based on the observed counts to the
lightcurves in the different energy channels and  re-computing
the CCF in 100 realisations for each burst. 
The
50th ranked value is  the mean lag and the 16th and 84th
ranked values represent $\pm1\sigma$.

\begin{figure} 
\begin{centering}
\epsscale{1}
\plotone{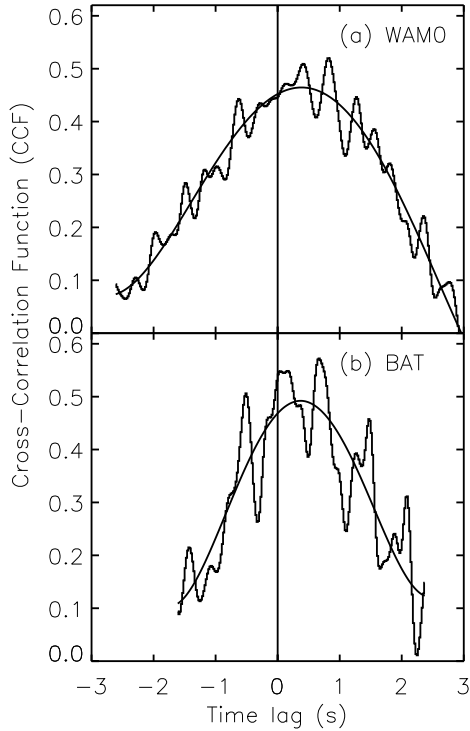}
  \caption{The cross-correlation data and fit 
  for a)
  the WAM0 data at 100\,ms between the 110-240 and 50-110\,keV energy bands
  and b)
   the  BAT data  at  100\,ms
   between the 50--100 and 25--50\,keV bands.  A 4$^{\rm th}$
           order polynomial fit to the data is shown. The vertical lines
           denote zero lag. GRB\,060505 is clearly inconsistent with zero
           lag.} \label{fig:ccfsWAM}
\end{centering}
\end{figure}

The spectral lag was determined between the 50--110 and 110--240\,keV ($\tau_{110-240,50-110}$)
energy bands for the $\emph{Suzaku}$ WAM detectors  over a range of temporal resolutions
(15.6, 31.2, 46.8, 62.4, 78 and 100\,ms).  The lightcurves were correlated from $-$4 to +4\,s and
the CCF was fit over a range of $\sim$5\,s. 
A lightcurve threshold of 10\% (30\%) is
applied, which means that only data with at least one-tenth (three-tenths) of the peak
count rate is used to calculate the lag, thus reducing the background. 
The spectral lag values obtained from WAM0 at the six  time resolutions specified above at 10\% threshold  agree within the uncertainties and the average value  is $0.36\pm{0.05}$\,s. 
Applying the 30\% threshold to the same lightcurves, increases the average value to 0.44$\pm{0.06}$\,s and all values are  within $1\sigma$ of those obtained at the 10\% threshold except at 60\,ms and 78\,ms, which are consistent at the  $2\sigma$ level.  
Above the 50\% threshold the results are unreliable 
and the lag is not accurately reproduced.
The burst is detected with lower significance in WAM3 and does not allow an accurate
determination of the lag. We add the signal from WAM3 to that of WAM0 to
test if this gives a consistent result. 
The results are consistent with WAM0 alone within $\sim1\sigma$, except at 100\,ms
resolution where the WAM0+3 lag is larger but is consistent at
$\sim3\sigma$ with the WAM0 results (10\% threshold). 
The average value obtained from the sum of  the WAM0 and WAM3 lightcurves
 is 0.42$\pm$0.05\,s and 0.47$^{+0.05}_{-0.06}$\,s at 10\% and 30\% thresholds respectively.
The cross correlation data and fit  for WAM0 at 100\,ms is presented in
Fig.~\ref{fig:ccfsWAM}\,a and is inconsistent with the negligible lag expected for
a  short burst. 
A precursor is evident in  the WAM data at $-$8\,s and including it in the lag analysis over a wider time range results in a consistent lag measurement of 0.47$\pm0.06$\,s. 
We note that precursors are not normally detected in short bursts.

The lag was also measured between the 25--50\,keV and 50--100\,keV energy
bands ($\tau_{50-100,25-50}$) at 100\,ms using the BAT data with the techniques outlined above  (Fig.~\ref{fig:ccfsWAM}\,b). The
lightcurve was too weak to determine the lag at finer time resolution. The
spectral lag value of 0.4$\pm$0.1\,s measured using the BAT data is
consistent within $1\,\sigma$ with that obtained from the WAM0 and WAM0+3.

In order to establish the robustness of our result, we determined the lag for 16 additional 
GRBs detected by both BAT and WAM, for which the  lightcurve data were sufficient for lag analysis in both instruments. 
The analysis was performed in a similar manner to GRB\,060505.   The derived lags ranged from $-$\,3\,ms to 0.94\,s in the WAM and 0 to 0.86\,s in the BAT.  
The lags  are compatible considering the differing instruments and off-axis angles, energy ranges and the spectra of the bursts, except in 2 cases where the BAT lag was significantly longer.  The sample  consisted of 12 long and 4 short bursts. 
Crucially,  the short bursts were always found to have negligible lag in both instruments.  This shows that 
our analysis is sensitive to short lags.  

\section{Discussion}

\subsection{Spectral lags}

\begin{figure}[t]
\epsscale{1}
\plotone{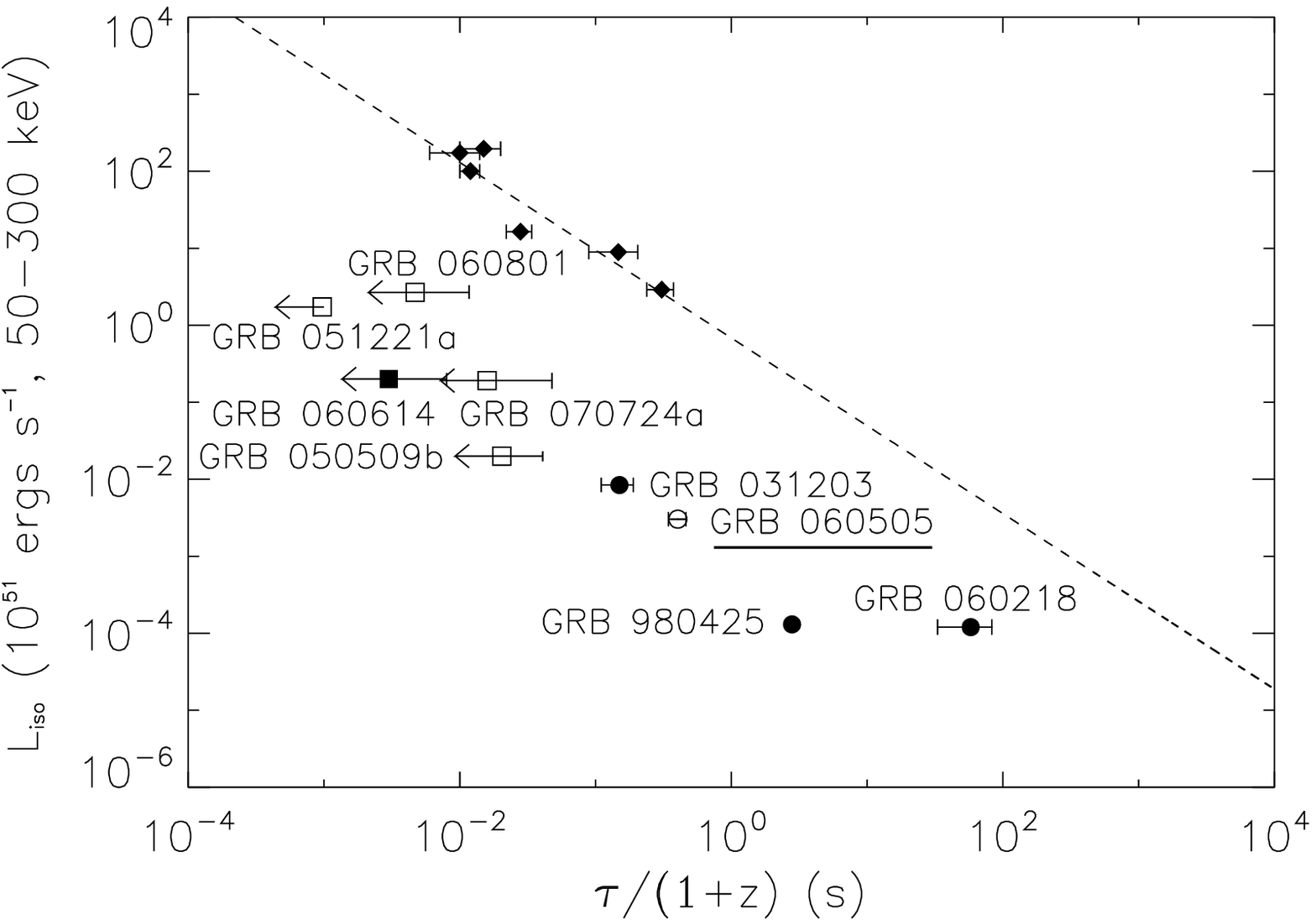}
 \caption{
     The lag-luminosity relation using data (diamonds) and fit from  \citet{2000ApJ...534..248N}.
 In addition, GRB\,060505 (open circle), GRB\,060614  \citep[filled square:][]{2006Natur.444.1044G}, short GRBs (open squares)
and 3 GRBs associated with SNe  (filled-circles) are included. The lag values are from the following:  GRB\,060218: \citet{2006ApJ...653L..81L}, 
GRB\,031203: \citet{Foley2007}, GRB\,980425: \citet{2002ApJ...579..386N}.  
       The spectral lag of GRB 060505 is significantly longer than those measured for short GRBs,            and it falls on the lag-luminosity plot in a position similar to that of some SN-GRBs.
                             The diamond and filled-circle lag values are determined between the 25--50 and 100--300\,keV  energy ranges. Lags for  GRB\,060614 and the short bursts are measured between the 15--25 and 50--100\,keV ranges. No $k$-correction is applied. }
           \label{fig:lag-lum}
\end{figure}

The spectral lags in GRBs have been discussed by many authors
\citep[e.g.][]{1997ApJ...486..928B,2000ApJ...534..248N,2000HEAD....5.3402N,2006ApJ...643..266N,2006ApJ...646.1086H,2007ApJS..169...62H,Foley2007}.

Using BATSE data, \citet{2002ApJ...579..386N} and
\citet{2006ApJ...643..266N} found that long duration GRBs had both
measurable and zero lags but that short GRBs had lags around zero.
\citet{2006ApJ...643..266N} calculated the lags of 260 short GRBs using
BATSE data and found that 90--95\% of the values were consistent with zero
and suggest that bursts with positive lag may result from contamination by the 
 long GRB class.
It was also argued that if short GRBs had lags proportionally as large as
long GRBs, such lags would be detectable, i.e.\ that this result was not
simply an effect of the duration of short bursts. This is not to say
that bursts with short lags are necessarily in the short GRB class. In the
sample of published lags of BATSE data by \citet{2007ApJS..169...62H}, 1427
bursts have $T_{\rm 90}\geq2$\,s and a measured lag ($\tau_{100-50,20-50}$). 
Of these bursts 214
have lags in the range from $-10$ to $+10$\,ms (79 with uncertainties of $\pm10$\,ms) and 348 have lags in the range from $-20$ to $+20$\,ms (217 with
uncertainties of $\pm20$\,ms), showing that there are many
long GRBs with very short lag. 
In summary, 
long bursts are expected to have predominantly positive lags ranging from zero to several seconds. 
Short GRBs have almost exclusively negligible lags. 
However, it is not possible to exclude that GRB\,060505 could be   an outlier: i.e. a short duration GRB  with a positive lag 
or due to a process which does not fit into the lag classification scheme.

There have been
difficulties in classifying a number of bursts  
and the lag has been used to discriminate 
in a
number of cases \citep[e.g.][]{2006astro.ph..5570D}. For example, GRB\,060912A has a T$_{\rm 90}$ of $\sim6$\,s and
 was initially thought to have occurred
in a nearby elliptical galaxy, however \citet{2007MNRAS.378.1439L} recently
found that it was more likely to come from a star forming galaxy at
$z=0.937$ and report a lag  ($\tau_{100-350,25-50}$) of $83\pm43$\,ms.
Various strategies have been proposed to distinguish bursts more effectively
than the duration alone
\citep[e.g.][]{2000ApJ...534..248N,2006astro.ph..5570D,2007ApJ...655L..25Z}.
However, none  have  seen widespread adoption.

\subsection{What was the progenitor of GRB\,060505?}

It was argued that GRB\,060505 was probably part of the tail of the short
burst population and connected to mergers of compact objects. 
At a redshift of $z$=0.089, GRB\,060505 has an isotropic peak
luminosity of $\sim9\times10^{47}\,\rm{erg}\,\rm{s}^{-1}$ (50-300\,keV). 
Having a low luminosity and relatively long lag of 0.36$\pm{0.05}$\,s, GRB\,060505 falls
below the lag-luminosity relation of~\citet{2000ApJ...534..248N}
as shown in Fig.~\ref{fig:lag-lum}. 
 The spectral lag of GRB 060505 is significantly longer than those measured for short GRBs and GRB\,060614  
and it falls on the lag-luminosity plot in a position similar to that of some (but not all) SN-GRBs (e.g. GRB\,031203). 

\citet{2007ApJ...662.1129O} argue that the simplest interpretation for GRB\,060505 
is that it is
related to a merger event rather than a short-lived massive star and point out
that the maximum allowable distance  of GRB\,060505 
from a star-forming knot is consistent with the shortest merger timescales.
\citet{2007astro.ph..3407T} claim that GRB060505 occurred in a star forming region of the host galaxy which   resembles long GRB host galaxies and argue for a massive star origin for this event.  It has
also been argued that the host galaxy of GRB\,060505 is more 
similar a short  burst  host  in terms of metallicity and ionisation state
\citep{2007ApJ...667L.121L}. 
However, their short GRB host region in the emission line ratio
diagram is based on only two burst host galaxies, one of which is the
GRB 050416A, which has photometric evidence for an associated SN
\citep{2007ApJ...661..982S} and is argued to be a long GRB due its spectral
softness and location on the Amati plot  \citep{2006ApJ...636L..73S}.
The host galaxy studies alone do not resolve the classification issue for   
 GRB\,060505.  
The optical luminosity at 12 hours in the source frame is similar to 
those of short GRB afterglows, but optical luminosity alone is also 
not a valid classification tool \citep{Kann2008}. 
In our opinion, the lag measurement suggests that this burst is similar to long GRBs implying a massive star progenitor, despite the lack of a SN detection. 

It has been argued that the absence of a SN signature in GRB\,060505  
is evidence of a new, quiet endpoint for some massive
stars \citep{2006Natur.444.1047F,2007astro.ph..3678W}.   
The existence of a SN was a feature of the early collapsar model. However,
the complete absence of a SN may be expected where the $^{56}$Ni does not 
have sufficient impetus to escape the black hole
\citep{2003ApJ...591..288H,2006ApJ...650.1028F}  or
in jet-induced explosions with narrow jets when the deposited energy is small
 \citep{2007astro.ph..2472N,2007ApJ...657L..77T}.
Progenitor stars with
relatively low angular momentum could also produce GRBs without supernovae
\citep{2003AIPC..662..202M}. 
 These seem  attractive
explanations at least for GRB\,060505. 
In the absence of a GRB explosion, the detection of such quiet-death massive stars, if they exist,
is a challenge for current instrumentation \citep[e.g.][]{2008arXiv0802.0456K}.

\acknowledgements
SMB acknowledges an EU 
  Marie Curie Fellowship in Framework 6.
The Dark Cosmology Centre is funded by the DNRF. DM acknowledges IDA for support.


\end{document}